\documentclass{webofc}

\usepackage[varg]{txfonts}
\usepackage{hyperref}
\usepackage{url}
\hypersetup{colorlinks=true,citecolor=blue,urlcolor=blue,linkcolor=blue}

\begin{document}

\title{Universal critical dynamics near the chiral phase transition and the QCD critical point}

\author{\firstname{Yunxin} \lastname{Ye}\inst{1}\fnsep\thanks{Speaker, \email{yunxin.ye@uni-jena.de}} \and
        \firstname{Johannes} \lastname{Roth}\inst{2} \fnsep\thanks{\email{johannes.v.roth@physik.uni-giessen.de}}
        \and
        \firstname{S\"oren} \lastname{Schlichting}\inst{3} 
        \and
        \firstname{Lorenz} \lastname{von Smekal}\inst{2,4}
}

\institute{Theoretisch-Physikalisches Institut, Abbe Center of Photonics, Friedrich-Schiller-Universit\"at Jena, Max-Wien-Platz 1, 07743 Jena, Germany \and
Institut f\"ur Theoretische Physik,
Justus-Liebig-Universit\"at, 
35392 Giessen, Germany \and
Fakult\"at f\"ur Physik, Universit\"at Bielefeld, D-33615 Bielefeld, Germany \and
Helmholtz Research Academy Hesse for FAIR (HFHF), Campus Giessen, 35392 Giessen, Germany}

\abstract{We use a novel real-time formulation of the functional renormalization group (FRG) 
for dynamical systems with reversible mode couplings to study Model G and H, which are the conjectured dynamic universality classes of the two-flavor chiral phase transition and the QCD critical point, respectively. 
We compute the dynamic critical exponent in both models in spatial dimensions $2<d<4$. We discuss qualitative commonalities and differences in the non-perturbative RG flow, such as weak scaling relations which hold in either case versus the characteristic strong scaling of Model G which is absent in Model H. For Model G, we also extract a novel dynamic scaling function that describes the universal momentum and temperature dependence of the diffusion coefficient of iso-(axial-)vector charges  in the symmetric phase.
}

\maketitle

\section{Introduction}
\label{sec:intro}

Understanding the real-time dynamics of hot and dense QCD matter in relativistic heavy-ion collisions from first principles is a challenging task. Near a second-order phase transition
the divergent correlation time due to critical slowing down entails that the dynamics of the long-wavelength modes becomes universal.
Corresponding \emph{dynamic} universality classes were classified by Halperin and Hohenberg \cite{RevModPhys.49.435}.

The chiral phase transition in the two-flavor chiral limit is expected to be second order and to fall into the $O(4)$ universality class as long as the axial $U(1)_A$ symmetry is not effectively restored at the critical temperature \cite{Pisarski:1983ms}. Since the physical up and down quark masses are much smaller than other typical QCD scales,
it is not unreasonable to expect dynamical effects from the $O(4)$ chiral phase transition as the fireball passes the chiral crossover region \cite{Florio:2021jlx}. An argument by Rajagopal and Wilczek suggests that the corresponding dynamic universality class is the one of Model G, i.e.,~a Heisenberg antiferromagnet \cite{Rajagopal:1992qz}.  

A true second-order phase transition in QCD with physical quark masses would be present in the form of the conjectured $Z_2$ critical point at finite baryon chemical potential. In this case, an argument by Son and Stephanov suggests that the corresponding dynamic universality class is the one of Model H, i.e.,~the liquid-gas critical point in a pure fluid \cite{Son:2004iv}.

For static critical phenomena, the functional renormalization group (FRG) can produce quantitative results for, e.g.,~critical exponents and amplitude ratios. 
On the other hand,  dynamic critical phenomena require a real-time formulation of the FRG.
While the real-time FRG has been successfully employed in the description of the purely relaxational Models A, B, \& C, Models G and H, on the other hand, are more subtle since they involve 
`reversible mode couplings'. Here, we summarize some results of our previous work \cite{Roth:2024rbi,Roth:2024hcu} and discuss how a real-time FRG flow which preserves the associated symmetries can be formulated.

\section{Real-time FRG approach to Model G/H}
\label{sec:dynUnivClasses}

We consider the Landau-Ginzburg-Wilson (LGW) free energy 
\begin{equation}
    F =\!\int \! d^d x \left\{ \frac{1}{2}(\partial^i \phi_a)(\partial^i \phi_a)+\frac{m^2}{2} \phi_a \phi_a + \frac{\lambda}{4!N}(\phi_a\phi_a)^2 \right\} 
    \label{eq:freeEnergyPhi}
\end{equation}
with $m^2 < 0$ to spontaneously break the $O(N)$ symmetry of the order parameter, and a positive quartic coupling $\lambda > 0$ for stability at large field values.

In Model G, the order parameter is not conserved, but couples reversibly to an antisymmetric tensor of charge densities $n_{ab}$, which are the generators of $O(N)$ transformations. Since their static fluctuations are non-critical, they enter the LGW free energy \eqref{eq:freeEnergyPhi} as $F \to F + \int d^d x \, n_{ab}n_{ab}/4\chi$.
The equations of motion of Model~G can be written as \cite{Rajagopal:1992qz}
\begin{subequations}
\begin{align}
    \frac{\partial \phi_a}{\partial t} &= -\Gamma^{\phi} \frac{\delta F}{\delta \phi_a}+\frac{g}{2}\,\{\phi_a,n_{bc}\} \, \frac{\delta F}{\delta n_{bc}}+\theta_a \,, \label{eomop} \\
    \frac{\partial n_{ab}}{\partial t} &= \gamma \vec{\nabla}^2 \frac{\delta F}{\delta n_{ab}}+g\,\{n_{ab},\phi_c\} \, \frac{\delta F}{\delta \phi_c}+ \frac{g}{2}\,\{n_{ab},n_{cd}\} \, \frac{\delta F}{\delta n_{cd}}+\zeta_{ab}  \, , \label{eomcc}
\end{align} \label{eq:eomsG}%
\end{subequations}
where $\theta_a$ and $\zeta_{ab}$ are Langevin noise terms whose variances set by the fluctuation-dissipation relation, and $\{\cdot,\cdot\}$ denotes the Poisson bracket, which is explicitly spelled out in Ref.~\cite{Roth:2024hcu}. 

In Model H, the order parameter has $N=1$ component, is conserved, and couples reversibly to the transverse part $\vec{j}$ of the momentum density. The equilibrium fluctuations of $\vec{j}$ are non-critical, which means that $\vec{j}$ appears in \eqref{eq:freeEnergyPhi} only quadratically,
$F \to F + \int d^dx \, \vec{j}^2/2\rho$,
where $\rho$ is the mass density. 
The equations of motion of Model~H are given by 
\begin{subequations}
\begin{align}
    \frac{\partial \phi}{\partial t}  &= \sigma \vec{\nabla}^2 \frac{\delta F}{\delta \phi} + g\{\phi,\vec{j}\} \cdot \frac{\delta F}{\delta \vec{j}} + \theta  \, ,\label{eomopH} \\
    \frac{\partial j_l}{\partial t} &= \mathcal{T}_{lm} \bigg[ \eta \vec{\nabla}^2 \frac{\delta F}{\delta j_m}+ g\{j_m,\phi\} \frac{\delta F}{\delta \phi} +  g\{j_m,j_n\}\frac{\delta F}{\delta j_n} \bigg]   + \xi_l \, , \label{eomojH}
\end{align}\label{eq:eomsH}%
\end{subequations}
where $\mathcal{T}$ is a transverse projector,
and $\theta$ and $\xi_l$ are again Langevin noise terms.

The basis for the real-time functional renormalization group is the Martin-Siggia-Rose (MSR) path-integral reformulation of the stochastic equations \eqref{eq:eomsG} and \eqref{eq:eomsH}. 
Importantly, the structure of the Poisson-bracket terms imply an extended temporal gauge symmetry of the MSR action \cite{Roth:2024rbi}, which is not present in the relaxational Models A, B \& C. This extended symmetry can be preserved during the FRG flow by introducing the bilinear regulator term at the level of the LGW free energy \eqref{eq:freeEnergyPhi}, rather than in the MSR action. While this makes no difference in the relaxational Models A, B \& C, the Poisson brackets in \eqref{eq:eomsG} and \eqref{eq:eomsH} entail that -- at the level of the MSR action -- the regulator couples to \emph{composite} response fields. Introducing the regulator in this way ensures 
that the flow of free energy is independent of dynamics, and that the
mode-coupling constant $g$ is protected from loop corrections.

\section{Results}

\begin{figure*}
    \centering
    \begin{minipage}{0.49\linewidth}
        \centering
        \begin{minipage}{0.65\linewidth}
            \centering
            \textsf{\normalsize \hspace{0.6cm} Model G}\\
            \includegraphics[width=0.95\linewidth]{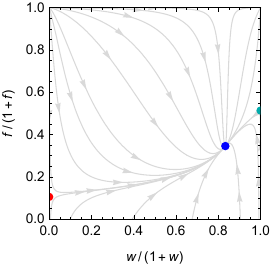}
        \end{minipage}
        \hfill
        \begin{minipage}{0.32\linewidth}
            \centering
            \phantom{\large Model G}
            \includegraphics[width=\linewidth]{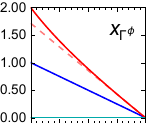}\\ \vspace{-0.04cm}
            \includegraphics[width=\linewidth]{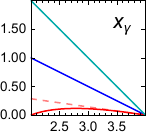}\\
        \textsf{\scriptsize spatial dimension \emph{d}}
        \end{minipage}  
    \end{minipage}
    \begin{minipage}{0.49\linewidth}
        \centering
        \begin{minipage}{0.65\linewidth}
            \centering
            \textsf{\normalsize \hspace{0.6cm} Model H}\\
            \includegraphics[width=0.95\linewidth]{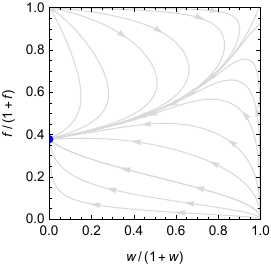}
        \end{minipage}
        \begin{minipage}{0.335\linewidth}
            \centering
            \phantom{\large Model H}
            \includegraphics[width=\linewidth]{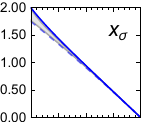}\\ \vspace{-0.10cm}
            \includegraphics[width=\linewidth]{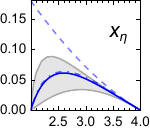}\\
        \textsf{\scriptsize spatial dimension \emph{d}}
        \end{minipage}
    \end{minipage}
    \caption{FRG flow diagrams of $3d$ Model G and H. Model~G admits one attractive strong-scaling fixed point (blue) at $(w^*,f^*) \approx (4.998,0.527)$, and two unstable weak-scaling fixed points (red and green) at $w^*=0,\infty$. Model~H admits one attractive weak-scaling fixed point (blue) at $w^* = 0$.  Critical exponents of the kinetic coefficients are plotted as a function of spatial dimension $2<d<4$ in matching colors. Here, solid lines indicate our FRG results with $\phi^4$-truncation, and dashed lines results from the perturbative $\epsilon$-expansion.
    Gray bands show FRG results with the extended local-potential approximation (LPA') of the free energy, with the upper and lower bounds corresponding to the field expansion points $\phi=0$ and the scale-dependent minimum of the effective potential. Figure taken from Ref.~\cite{Roth:2024hcu}.}
    \label{fig:ModGHFlow}
\end{figure*}

We truncate the FRG flow by promoting the couplings $m^2$ and $\lambda$ in LGW free energy \eqref{eq:freeEnergyPhi} to depend on the FRG scale (which we call `$\phi^4$-truncation'), as well as the kinetic coefficients $\Gamma^{\phi}$ and $\gamma$ in \eqref{eq:eomsG} and $\sigma$ and $\eta$ in \eqref{eq:eomsH}. Second-order phase transitions correspond to fixed points of the FRG flow. Suitable dimensionless combinations of the kinetic coefficients are given by 
\begin{align}
    w_G \equiv \chi \, \frac{\Gamma^{\phi}_k }{\gamma_k} \,, \hspace{0.4cm}  f_G \equiv \frac{d\,\Omega_d\, g^2 T}{(2\pi)^d} \, \frac{k^{d-4}}{ \Gamma_k^{\phi} \gamma_k} \,, \hspace{0.7cm}  w_H \equiv \rho \, \frac{\sigma_k   k^2}{\eta_k} \,, \hspace{0.4cm} 
    f_H \equiv \frac{d\,\Omega_d\, g^2 T}{(2\pi)^d} \, \frac{k^{d-4}}{\sigma_k \eta_k} \, .\label{definitionwfH}
\end{align}
Their FRG flow is shown in Fig.~\ref{fig:ModGHFlow}.
At the fixed points, the kinetic coefficients behave as power laws with the FRG scale, $\Gamma_k^{\phi} \sim k^{-x_{\Gamma^{\phi}}}$, $\gamma_k \sim k^{-x_{\gamma}}$, $\sigma_k \sim k^{-x_{\sigma}}$ and $\eta_k \sim k^{-x_{\eta}}$.

Qualitatively, the finite fixed-point value of $0<f_{G/H}^*<\infty$ implies the scaling relation $x_{\Gamma^{\phi}}+x_{\gamma}=4-d-\eta_{\perp}$ in Model~G, and $x_{\sigma} + x_{\eta} = 4-d-\eta_{\perp}$ in Model~H (here including the anomalous dimension $\eta_{\perp}$ of the order parameter). On the other hand, the fixed-point value $w_G^*$ is finite only at the strong-scaling fixed point of Model~G, where it implies strong-scaling behavior with 
$z_{\phi}=z_{n}=d/2$. This relation is absent at the other fixed points, which either have $w^*_{G/H}=0$ or $w^*_{G/H}=\infty$. In the phenomenologically relevant case of $d=3$ spatial dimensions we obtain the values $x_{\sigma} \approx 0.949$ and $x_{\eta} \approx 0.051$ in Model~H, which correspond to the dynamic critical exponent $z_{\phi}=d+x_{\eta} \approx 3.051$. We observe for $d=2$ that the critical exponent of the shear viscosity vanishes, $x_{\eta} = 0$.

\begin{figure}
    \centering
    \sidecaption
    \includegraphics[width=0.5\textwidth,clip]{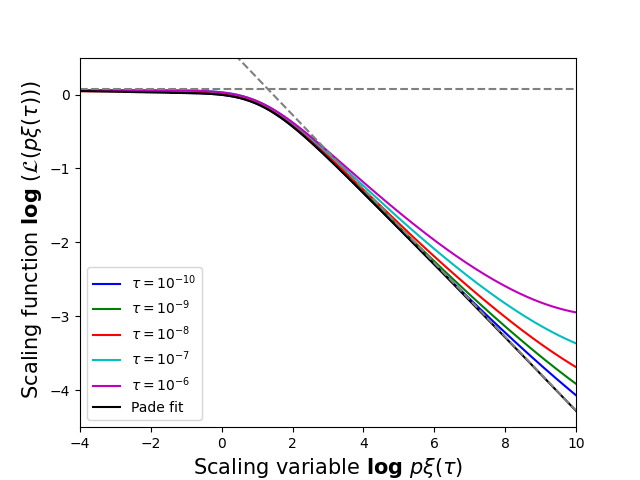}
    \caption{By tuning the temperature close to the critical temperature from above, and rescaling the resulting diffusion coefficient $D_{n}(\vec{p})$ at $k=0$ according to \eqref{eq:Dn}, we find a scaling function $\mathcal{L}(x)$ which 
    only depends on the scaling variable $x\equiv p\xi(\tau)$. Figure taken from Ref.~\cite{Roth:2024rbi}.}
    \label{fig:DnScalFnc}
\end{figure}

In Ref.~\cite{Roth:2024rbi}, we have supplemented our truncation of Model G by momentum dependence of iso-(axial-)vector charge mobility $\gamma_k \to \gamma_k(\vec{p})$. The dynamic scaling hypothesis for the associated diffusion coefficient $D_n(\vec{p}) \equiv \gamma(\vec{p})/\chi$ implies the scaling form (at $k=0$)
\begin{equation}
    D_n(\vec{p},\tau) = D_n^+ \tau^{-\nu x_{\gamma}} \mathcal{L}(p\xi(\tau)) \label{eq:Dn}
\end{equation}
for its momentum and temperature dependence in the scaling regime. Here, $\tau\equiv (T-T_c)/T_c$ is the reduced temperature, and $\xi(\tau) = 1/m_{k=0}$ is the correlation length. As discussed above, we have $x_{\gamma} = 2-z_n = 2-d/2$ at the strong-scaling fixed point. By tuning the temperature close to the critical temperature from above, and rescaling the resulting $D_n(\vec{p},\tau)$ according to \eqref{eq:Dn}, the curves collapse onto a scaling function $\mathcal{L}(x)$, which is shown in Fig.~\ref{fig:DnScalFnc}.

\section{Outlook}

The present formulation of the real-time FRG 
lays the foundation for more sophisticated truncation schemes in the future. For example, by including momentum dependence in the kinetic coefficients of Model~H one can compute scaling functions which go beyond the commonly employed Kawasaki approximation. On the other hand, by including external fields in Model~G (which correspond to finite current quark masses in QCD), one can study the influence of critical dynamics on hot QCD matter at the physical point within a low-energy effective theory. Besides dynamic critical phenomena, another application of the present formalism is fluctuating hydrodynamics, where the real-time FRG can be used to study the influence of fluctuations on transport coefficients non-perturbatively.  \\ \vspace{-0.3cm}

{\footnotesize
\noindent This work was supported by the Deutsche Forschungsgemeinschaft (DFG, German Research Foundation) through the CRC-TR 211 ‘Strong-interaction matter under extreme conditions’-project number 315477589 – TRR 211. JVR is supported by the Studienstiftung des deutschen Volkes.}

 \vspace{-0.3cm}


\bibliography{refs}
\end{document}